\documentclass[floatfix,showpacs,amsmath,amssymb,letterpaper,groupaddresses,superscriptaddress]{article}
\usepackage{graphicx}
\usepackage{latexsym}
\usepackage{amsmath}
\usepackage{amssymb}
\usepackage{epsfig}
\usepackage{graphicx}
\usepackage{graphicx}
\usepackage{amssymb,amsmath,times}
\usepackage{hyperref}
\usepackage{color}

\def\a{\alpha}
\def\b{\beta}

\def\be{\begin{equation}}
\def\ee{\end{equation}}
\def\ba{\begin{eqnarray}}
\def\ea{\end{eqnarray}}
\def\la{\langle}
\def\ra{\rangle}
\def\a{\alpha}

\def\b{\beta}

\def\h{\hskip 1cm}

\def\bz{\mathbf{0}}\def\bo{\mathbf{1}}\def\bt{\mathbf{2}}
\def\bk{\mathbf{k}}
\def\bN{\mathbf{N}}\def\bth{\mathbf{3}}
\def\bfo{\mathbf{4}}\def\bfi{\mathbf{5}}
\def\bsi{\mathbf{6}}

\usepackage{epsfig}
\usepackage{graphicx}
\begin{document}
\begin{center}

{\Large \bf Noise effects in perfect transmission of quantum states}\\
\vspace{1cm} Fabio
Benatti$^{a,b}$\footnote{email:benatti@ts.infn.it}\h Roberto
Floreanini$^b$\footnote{email: roberto.floreanini@ts.infn.it}\h
Vahid Karimipour$^c$\footnote{email:vahid@.sharif.edu}
\vspace{5mm}

\vspace{1cm} ${}^a$Dipartimento di Fisica, Universit\`a degli
Studi di Trieste, I-34151 Trieste, Italy,\\

\vspace{1cm} ${}^b$Istituto Nazionale di Fisica Nucleare, Sezione
di Trieste, I-34151 Trieste, Italy\\

\vspace{1cm} ${}^c$Department of Physics, Sharif University of
Technology, P.O. Box 11155-9161, Tehran, Iran.
\end{center}
\vskip 3cm

\begin{abstract}
A recent scheme for perfect transmission of quantum states
through quasi-one dimensional chains requires application of
global control at regular intervals of time. We study the effect
of stochastic noise in this control and find that the scheme is
robust for reasonable values of disorder.  Both un-correlated and
correlated noise in the external control are studied and it is
remarkably found that the efficiency of the protocol is much
higher in presence of correlated noise.
\end{abstract}

\section{Introduction}

Since the early developments in the theory of quantum information, the task of coherently
transferring quantum states through long and short distance communication lines has
been of great importance. While photons are the ideal carriers
of quantum information over long distances, it has become evident
that the best possible method for transferring quantum
information over short distances is to exploit the dynamics of many body
systems, specifically of regular arrays of qubits constituting suitable spin chains.
In this framework, an arbitrary qubit state is first coupled to
the array and then carried to destination by the natural dynamics of the whole system,
where it can be extracted with certain fidelity.
This idea was first introduced in \cite{Bose}, where it was shown that the natural dynamics of a
Heisenberg ferromagnetic spin chain can achieve high-fidelity transfer of
qubits over distances as long as 80 lattice units.%
%
%\footnote{For additional information, see \cite{Bose2} and references therein.}
%
In contrast to
this traditional "passive'' protocol, different approaches soon emerged. One idea was
to engineer the couplings between the various spins in the chain in such a way that
states are transferred with perfect \cite{Christandl}-\cite{Markiewicz} or with
arbitrary high fidelity \cite{Shi}-\cite{Yao}; in addition, some minimal external control on the chain
dynamics was also introduced in order to
achieve similar results \cite{Burgarth1}-\cite{Banchi}. \\

Quite recently it was shown \cite{Pemberton} that particular types of quasi-one dimensional uniformly coupled chains, can
achieve perfect state transfer, provided that the natural dynamics of the chain be supplemented with some global control
pulses at regular intervals of time. Then it was shown in \cite{Pemberton} and using  a different geometry in \cite{KSA},
 that perfect routing of states in higher dimensional networks, from any point to any other point, is also possible.
 The advantage of these schemes was that they allowed simultaneous routing of multiple states and also the possibility of
 overcoming some of the restrictions of the previous protocols, notably introducing some robustness to local imperfections
 in the network. It is important to note that the introduction of external control is not in contrast with the spirit of
 quantum state transfer through the natural dynamics of the chain, as long as the external control is global and does not
 address individual qubits in the network.  \\

While these schemes are to some extent robust against localized imperfections in couplings, i.e. by routing around
known defects in the network, the new element of global control inherent in these kinds of schemes brings about the
question of their robustness to inaccuracies and imperfections of external control. We can ask to what extent the
 fidelities of these schemes is affected by imprecision in the timings and the direction of the applied fields in the global pulses, which are
 necessary for perfect routing of states.  \\

It is on these novel stochastic disturbances that we shall focus in the present work.  Specifically we focus on the
quasi-one dimensional chain of \cite{Pemberton}, which is the basic element in higher dimensional uniformly coupled
networks. We show that for a moderate value of the noise in global control, one can still achieve a high value of fidelity.
 Interestingly, we find that the transfer protocol appears to be less affected by correlated noise, as compared
to un-correlated one; this result may be of great interest in the
actual realization of realistic spin chain channels, since it
suggests that externally induced time-correlations
may protect the efficiency of the spin transmission lines. \\

Although in general such noises affect the dynamics of the full
chain in an analytically intractable way, we show that the
fidelity of the state transfer can be exactly determined and
analyzed. More specifically, we are able to compute the fidelity
of the protocol, when the direction of the applied field and the
timings of the pulses are not precisely tuned because of the
presence
of external stochastic noise. \\

The structure of the paper is as follows. In the next section we
give a brief account of perfect state transfer in the quasi-one
dimensional chain of \cite{Pemberton}. Section 3 deals with the
general formalism that allows to incorporate in the protocol the
effects of the presence of disorder in the pulses. Section 4 is
instead devoted to the actual computation of the fidelity both in
presence of un-correlated disturbances, and in the more
interesting case of correlated noise. The final section contains
a brief discussion and outlook.

\section{Perfect state transfer in a uniformly coupled quasi-one dimensional chain}
The prototype of many-body system that has been used in many
protocols is the $XY$ spin chain, consisting of a linear array of
$N$ sites, to each of which a spin-1/2 operator with Cartesian
components $X$, $Y$, $Z$ is attached. The dynamics is then
described by a Hamiltonian of the form
\begin{equation}
\label{proto}
 H=\frac{1}{2}\sum_{m,n} J_{m,n}(X_m X_n+Y_m Y_n)\ ,
\end{equation}
where the sum is over the various bonds in the array
(see Fig.(1) for the specific example discussed in the following).
This type of interaction preserves the total component of the spin along the $z$-direction
\begin{equation}
\label{conserve}
\Big[H,\sum_m Z_m\Big]=\, 0\ ,
\end{equation}
and moreover does not evolve the uniform background state where all
the spin projections along the $z$ direction are up,
conventionally called the vacuum state $|\mathbf{0}\rangle$: $H|\bz\ra=0$.
Further, let us denote with $|\mathbf{i}\rangle$ the single excitation state corresponding to the situation
in which all spins are up except the one at site $i$ which is down; the collection
$\{\,|\mathbf{i}\rangle\,\}$, $i=1,2,\ldots, N$, of these $N$ states,
form a basis in the single excitation subspace of the
total Hilbert space of the system.

In order to transfer an arbitrary qubit state
$\begin{pmatrix}\alpha\cr\beta\end{pmatrix}$
from site $1$ to site $N$, one may apply the following simple protocol.
First, embed the qubit into the initial lattice state $\a|\bz\ra+\b|\bo\ra$.
Then, let it evolve according to the natural chain dynamics through the lattice until it becomes,
at a suitable instant of time, the state $\a|\bz\ra+\b|\bN\ra$; from this state, one can then extract
the original qubit from site $N$.

As mentioned before, one way to achieve this perfect state transfer is to carefully engineer
the coupling constants; this has been discussed in \cite{Christandl}, where it was shown that a linear $XY$ chain
with local couplings of the form $J_{n,m}=\sqrt{n(N-n)}\,\delta_{n+1,m}$ can indeed perfectly transfer
a qubit to the end of the chain, at the specific time $t=\pi$.
It is interesting to note that for $N=2$ and $N=3$, the couplings will be uniform and indeed it has been shown that these are the only uniformly coupled chains which can achieve perfect state transfer.

The specific chain analyzed in \cite{Pemberton} is shown in
Fig.(1). The chain is uniformly coupled in the sense that all the
couplings have the same modulus. The presence of $-1$ couplings
allows this chain to be broken up into a direct sum of sub-chains
with just two and three sites, achieving perfect state transfer.
To see this, consider the Hamiltonian pertaining to this chain;
with reference to the labeling of the sites of Fig.(1), it can be
expressed as
\begin{equation}
\label{Hamiltonian0a}
H=\Big(|\bo\ra\la \bt|+|\bo\ra\la \bth|+h.c.\Big)+\Big(|\bt\ra\la \bfo|-|\bth\ra\la \bfo|+h.c.\Big)
+\Big(|\bfo\ra\la \bfi|+|\bfo\ra\la\bsi|+ h.c.\Big)\,+\,\ldots
\end{equation}
which can be conveniently re-written as
\begin{equation}
\label{Hamiltonian0b}
H=\sqrt{2}\Bigg[\Big(|\bo\ra\la (\bt,\bth)_+|+h.c.\Big)+
\Big(|(\bt,\bth)_-\ra\la \bfo|+|\bfo\ra\la(\bfi,\bsi)_+|+h.c.\Big)+\ldots\Bigg]\ ,
\end{equation}
where
$$
|(\bk,\bk+\bo)_\pm\ra=\frac{|\bk\ra\pm|\bk+\bo\ra}{\sqrt{2}}\quad  .
$$

\begin{figure}[t]
\centering
\includegraphics[width=12cm,height=7cm,angle=0]{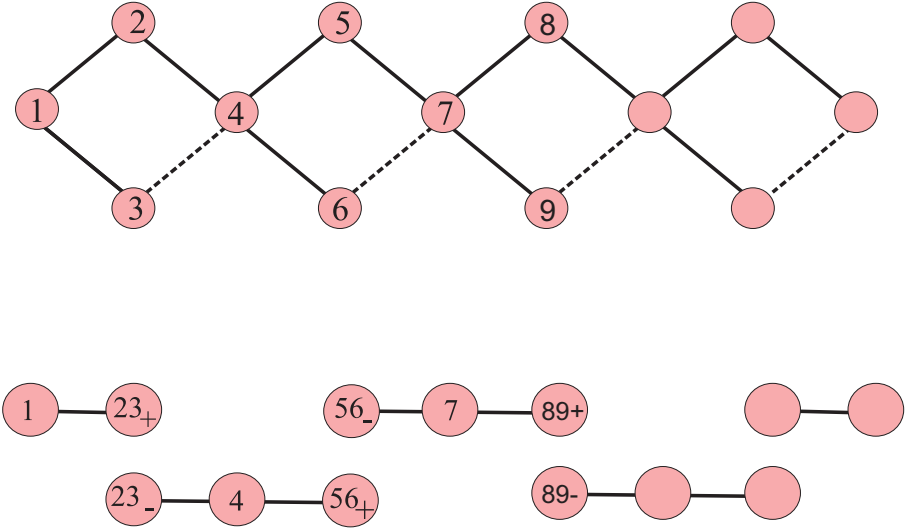}
\caption{(color online)\small Above: the quasi-one-dimensional chain introduced in \cite{Pemberton};
solid lines indicate bonds with coupling equal to 1, dashed lines those with coupling equal to -1.
Below: its equivalent description in terms of two- and three-site perfect state transfer chains.}
\label{fig1}
\end{figure}

The orthogonality of the states $|(\bk,\bk+\bo)_\pm\ra$ shows that the diamond-shaped
lattice can be studied as a chain formed by two-site and three-site elementary components as shown in Fig.(1);
we will refer to the latter as ``virtual chain'' and work with it from now on:
it consists of $K=(N-4)/3$ three-site sub-chains plus an initial and a final two-site sub-chain.
A convenient basis, in the single-excitation Hilbert sub-space is then given by
$\{|i\ra\}$, $i=1,2,\ldots,N$, made of the following orthonormal vectors
\begin{eqnarray}
\label{ONB}
&&\hskip -1cm|1\rangle\equiv|\bo\ra\ ,\quad |2\ra\equiv|(\bt,\bth)_+\ra\ ,\quad |3\ra\equiv|(\bt,\bth)_-\ra\ ,\quad
|4\ra\equiv|\bfo\ra\ ,\ldots\\
\nonumber
&&\hskip 4cm \ldots\ |N-1\ra\equiv|(\bN-\bo,\bN)_-\ra\ ,\quad |N\ra\equiv|\bN\ra\ .
\end{eqnarray}
Note that the vector states $|3i+1\ra$, $i=0,1,\ldots,(N-1)/3$,
correspond to single spin down at the site $3i+1$, like in the
diamond-shape representation of the chain; however, unlike in
that one, the other vectors represent a single-excitation being
in a superposition of a spin down at one site and  down at the
subsequent one. {\it From now on we shall work within the
representation based on the orthogonal states non-bold faced
states in (\ref{ONB}) corresponding to the disjoint union of
small chains.}\\

As mentioned before, the time-evolution operator $U_t=\exp(-itH)$ maps
the single-excitation sub-space into itself. In the representation corresponding to the above orthonormal basis,
it can be split into the following orthogonal sum of blocks
\be
\label{blockun}
U_t = \begin{pmatrix}
U_t^{(1)}&&&\cr
& U_t^{(2)}&&&\cr
&&U_t^{(3)}&&\cr
&&&\ddots&\cr
&&&& U_t^{(K+2)}\end{pmatrix}\ .
\ee
The unitary blocks $U^{(1)}_{t}$ and $U^{(K+2)}_t$, corresponding to the initial and final two-site
sub-chains, can be represented as
\be
\label{U2}
U^{(1)}_{t}=U^{(K+2)}_t=
{\rm e}^{-it\sqrt{2}\sigma_x}=\begin{pmatrix}\cos t\sqrt{2}&-i\sin t\sqrt{2}\cr -i\sin t\sqrt{2}&\cos t\sqrt{2}\end{pmatrix}\ ,
\ee
by means of the Pauli matrix $\sigma_x=\begin{pmatrix}0&1\cr1&0\end{pmatrix}$, while
the remaining ones can be represented as
\be
\label{U3}
U^{(j)}_t={\rm e}^{-i2t\,S_x}=\frac{1}{2}
\begin{pmatrix}1+\cos 2t&-i\sqrt{2}\sin 2t&-1+\cos 2t\cr
-i\sqrt{2}\sin 2t&2\cos 2t&-i\sqrt{2}\sin 2t\cr
-1+\cos 2t&-i\sqrt{2}\sin 2t&1+\cos 2t\end{pmatrix}\ , \quad j=2,\ 3,\ldots,K+1\ ,
\ee
in terms of the $x$ component of a spin-1 operator:
\be
\label{Sx}
S_x=\frac{1}{\sqrt{2}}\begin{pmatrix}0&1&0\cr1&0&1\cr0&1&0\end{pmatrix}\ .
\ee
The use of $\sigma_x$ and $S_x$ as generators of the blocks of the unitary time-evolution matrix $U_t$ follows from
the expression~(\ref{Hamiltonian0b}) for the Hamiltonian, whereby $\sigma_x$ represents the first term in the sum, $S_x$
the second one and so on. One easily sees that, for $t=\pi/(2\sqrt{2})$, respectively $t=\pi/2$, the unitaries in (\ref{U2}), respectively (\ref{U3}),  realize prefect state transfers over the elementary two-, respectively three-site sub-chains.

Besides the unitary evolution given by $U_t$, the transfer protocol introduced in \cite{Pemberton} involves
instantaneous strong pulses $P$ acting globally on the lower sites of the diamond chain in Fig.(1).
The explicit form of the operator $P$ is most simply given in the language of the virtual chain
and in the basis (\ref{ONB}):
\be
P=\begin{pmatrix}
1&&&&&&\cr
&\sigma_x&&&&&\cr
&&1&&&&\cr
&&&\sigma_x&&&\cr
&&&&\ddots&&\cr
&&&&&\sigma_x&\cr
&&&&&&1\end{pmatrix}\ ,
\label{pulse}
\ee
where $\sigma_x$ appears $K+1$ times, each one of them coupling $\vert i\rangle$ and $\vert i+1\rangle$ for $i=3n+2$ with $n=0,1,\ldots,K$.
Thus, except for the initial and final sites of the virtual chain,
the pulse $P$ transfers any single excitation at an end point of a given
sub-chain to that of the contiguous one. The basic idea of the transfer protocol of \cite{Pemberton}
is to use the unitaries $U_t$ to transfer
a single excitation along the sub-chains, while using $P$ to make it jump from one
sub-chain to the next. In practice,
the composite dynamics of the system, starting from the initial time $t=\,0$ to the
final time $t=t_f$, is then described by the unitary operator
\be
\label{UTE}
\mathcal{U}_{t_f}=U_{t_f-t_{K+1}}\,P\,U_{t_{K+1}-t_K}\,P\,\cdots\,U_{t_2-t_1}\,P\,U_{t_1}
\ .
\ee
Suppose the times $t_1$, $t_2$, \ldots, $t_{K+1}$ at which the pulses act are chosen as
\be
\label{time-int}
t_1=\frac{\pi}{2\sqrt{2}}\ ,\qquad  t_{j+1}=\frac{\pi}{2\sqrt{2}}+j\,\frac{\pi}{2}\ ,\quad j=1,2,\ldots,K\ .
\ee
Then, from the explicit form that (\ref{blockun}), (\ref{U2}) and (\ref{U3}) take in this case,
one easily sees that the initial state
$\alpha |0\rangle+ \beta |1\rangle$, where $|0\rangle$ denotes the ground state,
will be perfectly transformed into
$\alpha |0\rangle+ \beta |N\rangle$ at the final time
\be
\label{finaltime}
t_f=\frac{\pi}{\sqrt{2}}+\frac{\pi}{2}K\ .
\ee
From this final state, the single qubit $\begin{pmatrix}\alpha\cr\beta\end{pmatrix}$ can be extracted from site $N$,
thus realizing its perfect transfer from the beginning to the end of the chain.

\section{Noise in the protocol}

As mentioned in the introduction, in order to transfer a generic qubit state
\be
\label{initialstate}
|\psi\ra=\begin{pmatrix}\a\cr\b\end{pmatrix}\ ,\quad \a=\cos\frac{\theta}{2}\ ,\
\b=\sin\frac{\theta}{2}e^{-i\phi}\ , \quad 0\leq\theta\leq\pi\ ,\ 0\leq\phi\leq2\pi\ ,
\ee
along the chain, one first embeds it into the left hand site of the chain itself as the state
\be
\label{emb}
|\Psi\ra=\a|0\ra+\b|1\ra\ ,
\ee
where $\{|i\ra\}_{j=1}^N$ are the basis vectors in (\ref{ONB}) and $|0\ra$ is again the ground state.
Under the unitary action of the time-evolution (\ref{UTE}), $|\Psi\ra$ is transformed into $|\Psi_{t_f}\ra=\mathcal{U}_{t_f}|\Psi\ra$; since $H|0\ra=0$,
\be
\label{stateattf}
|\Psi_{t_f}\ra=\a|0\ra+\b\left(\sum_{i=1}^N\psi_i(t_f)\,|i\ra\right)\ .
\ee
The protocol purpose is to use the chain dynamics to transfer the state $|\psi\ra$ from site $1$ to site $N$; the state of the qubit at site $N$ is obtained by performing the partial trace ($tr_N$) over single-excitation states involving all sites but the $N$-th one of the projection
$|\Psi_{t_f}\ra\la\Psi_{t_f}|$; this gives a $2\times 2$ density matrix
\be
\label{densityN}
\rho_N(t_f)=tr_N|\Psi(t_f)\ra\la \Psi(t_f)|=\begin{pmatrix}
|\a|^2+|\b|^2(1-|\psi_N(t_f)|^2)& \a\b^*\,{\psi_N(t_f)}^*\cr  \a^*\b\,{\psi_N(t_f)}& |\b|^2|\psi_N(t_f)|^2
\end{pmatrix}\ .
\ee

The robustness of the transmission along the chain can be
measured by computing the fidelity of the final mixed state at
the site $N$ with respect to the initial pure one embedded at
site $1$; it is given by \be \label{fidelity}
F_\psi=\la\psi|\rho_N(t_f)|\psi\ra=|\a|^2+2\,|\a|^2|\b|^2\,Re(\psi_N(t_f))+|\b|^2(|\b|^2-|\a|^2)|\psi_N(t_f)|^2\
. \ee Note that it involves only the last coefficient
$\psi_N(t_f)$ in the expansion (\ref{stateattf}). Its average
over all input states, given by \cite{Bose} \be \label{avfid}
F:=\frac{1}{4\pi}\int_0^{2\pi}{\rm d}\phi\int_0^\pi{\rm
d}\theta\,\sin\theta\,F_\psi
=\frac{1}{2}+\frac{1}{6}|\psi_N(t_f)|^2+\frac{1}{3}Re(\psi_N(t_f))\
, \ee is a measure of the overall robustness of the transmission
protocol.

The fidelity is maximal, that is $F_\psi=1$, in the case of the
ideal perfect protocol embodied by the unitary evolution
$\mathcal{U}_{t_f}$ in (\ref{UTE}) with times as in
(\ref{time-int}) and pulses as in (\ref{pulse}). The success of
the perfect state transfer depends on the precise control over
the timing of the pulses and the direction of their field; in
particular, the pulse should be exactly of the form (\ref{pulse})
to perfectly transfer the state from one sub-chain to the next.
Also the timing of these pulses should be precise and synchronous
with the timing required for perfect state transfer within the
sub-chains. A disorder in either the field direction or timing of
the pulses may drastically reduce the fidelity of the final state.

In the following, we shall consider the case where external noise affects the protocol by modifying precisely these times and pulses.
In practice, instead of those in (\ref{time-int}),
we shall consider modified times of the form
(as before, the initial time is set conventionally to zero):
\begin{eqnarray}
\nonumber
&&t_1=\frac{\pi}{2\sqrt{2}}\,+\,\tau_1\ ,\\
\label{noise1}
&&t_{j+1}=\frac{\pi}{2\sqrt{2}}+j\,\frac{\pi}{2}\,+\,\tau_{j+1}\ ,\qquad j=1,2,\ldots,K\ ,\\
\nonumber
&&t_f=\frac{\pi}{\sqrt{2}}+\frac{\pi}{2}K\,+\,\tau_{K+2}\ ,
\end{eqnarray}
where $\tau=\{\tau_i\}_{i=1}^{K+2}$ is a stochastic process with random variables $\tau_i$ distributed according to joint probabilities ${\cal P}_{time}(\tau)={\cal P}_{time}(\tau_{K+2},\tau_{K+1},\ldots,\tau_1)$.
The stochastic process may be stationary or not, correlated or not and the stochastic variables may take real values in a discrete or continuous state space. For sake of simplicity, we shall only suppose them to have zero mean-values.

Analogously, we shall consider noisy pulses $P(\theta_i)$ that, while keeping the block form (\ref{pulse}), will
no longer consist of local pulses represented by $\sigma_x$, but by
\be
P(\theta_i):=\begin{pmatrix}i\sin\theta_i & \cos\theta_i \cr
\cos\theta_i & i\sin\theta_i \end{pmatrix}\ ,
\ee
that reduces to $\sigma_x$ in the limit $\theta_i=\,0$;
here, $\theta=\{\theta_i\}_{i=1}^{K+1}$ is also a stochastic process with probability distribution
${\cal P}_{pulse}(\theta)={\cal P}_{pulse}(\theta_{K+1},\theta_K,\ldots,\theta_1)$ consisting of real stochastic variables $\theta_j$ with zero mean values.

We shall collect the two stochastic processes into a single one, $\mu=(\tau,\theta)$,
with joint probabilities ${\cal P}(\mu)$, that may even account for possible correlations between them.
Then, in presence of such kind of noise, the unitary evolution (\ref{UTE}) will be replaced by
\be
\label{UTEnoisy}
\mathcal{U}^{(\mu)}_{t_f}=U(\tau_{K+2}-\tau_{K+1})P(\theta_{K+1})U(\tau_{K+1}-\tau_K)\cdots U(\tau_2-\tau_1)P(\theta_1)U(\tau_1)\ ,
\ee
where the dependence on the stochastic variables in the various contributions is explicitly shown.
Because of its very construction, despite the presence of a perturbing noise, $\mathcal{U}^{(\mu)}_{t_f}$ will map the single-excitation sub-space into itself.
Thus, for each realization of the noise, the initial state (\ref{emb}) will be mapped  into another single-excitation state
\be
\label{noisystate}
|\Psi^{(\mu)}_{t_f}\ra=\mathcal{U}^{(\mu)}_{t_f}|\Psi\ra=\a\,|0\ra\,+\,\b\,\sum_{i=1}^N\psi^{(\mu)}_i(t_f)\,|i\ra\ ,
\ee
with a reduced density matrix at the $N$-th site,
$\rho^{(\mu)}_N(t_f)=tr_{\hat{N}}|\Psi^{(\mu)}(t_f)\ra\la \Psi^{(\mu)}(t_f)|$,
given again by the matrix in (\ref{densityN}), but with
$\psi_N(t_f)$ replaced by $\psi^{(\mu)}_N(t_f)$,
and similarly for the corresponding averaged fidelity $F^{(\mu)}$ (see~(\ref{avfid})).

However, a physically meaningful state for the system can only be obtained by averaging over all realizations of the  noise; the density matrix representing the state of the chain at the final time $t_f$ is therefore given by
\be
\label{averstate}
\rho(t_f)=\la\rho^{(\mu)}(t_f)\ra\equiv\int{\rm d}\mu\ {\cal P}(\mu)\,|\Psi^{(\mu)}_{t_f}\ra\la\Psi^{(\mu)}_{t_f}|\ ,
\ee
where the integration region is given by the
space spanned by the values that the stochastic variables $(\tau, \theta)$ can take.
Then, the reduced $N$-site state at time $t_f$ will become
\be
\label{noisydensity}
\rho_N(t_f)=tr_N\Big(\rho(t_f)\Big)=\Big\langle tr_N\Big(\rho^\mu(t_f)\Big)\Big\rangle\ ,
\ee
so that the fidelity averaged over the disorder and over all initial states will be given by
\be
\label{avvfid}
\la F\ra=\int{\rm d}\mu\ {\cal P}(\mu)\, F^{(\mu)} =\frac{1}{2}
+\frac{1}{6}\Big\langle |\psi^{(\mu)}_N(t_f)|^2\Big\rangle
+\frac{1}{3}\Big\langle Re(\psi^{(\mu)}_N(t_f))\Big\rangle\ .
\ee

\section{Fidelity in the presence of noise}

We are now faced with the task of computing the fidelity $\la F\ra$ of the state
(\ref{noisystate}) with respect to the initial state to be transferred.
For this type of dynamics, it is obviously impossible to determine the explicit form
of the final state given any arbitrary initial state.
Nevertheless, as explicitly shown in (\ref{avvfid}), in order to evaluate $\la F\ra$, only the coefficient
$\psi^{(\mu)}_N(t_f)$ in the expansion (\ref{noisystate}) is
really needed.

As already stressed, the crucial observation is that, in order to determine $\psi^{(\mu)}_N(t_f)$, one needs just
follow the change of the last non-zero entry of the vector~$|1\ra$
under the sequential action of operators of the form (\ref{U2}) and (\ref{U3}).
In doing so, we shall explicitly write only those components of the transformed vector
affected by the various unitary blocks.
Let us then consider the initial state $|1\ra=(1,0,0\ldots,0)^T$; the action of
$$
U^{(1)}(\tau_1)=\begin{pmatrix}
-\sin(\sqrt{2}\tau_1)&-i\cos(\sqrt{2}\tau_1)\cr -i\cos(\sqrt{2}\tau_1) &-\sin(\sqrt{2}\tau_1)
\end{pmatrix}
$$
transforms it into
$\displaystyle
\begin{pmatrix}-\sin(\sqrt{2}\tau_1)\cr -i\cos(\sqrt{2}\tau_1) \end{pmatrix}$.
Then, the first pulse $P(\theta_1)$ maps the relevant two-component vector $\displaystyle
\begin{pmatrix}-i\cos(\sqrt{2}\tau_1)\cr 0 \end{pmatrix}$ to
$\displaystyle\begin{pmatrix}
\sin\theta_1\cos(\sqrt{2}\tau_1)\cr -i\cos\theta_1\cos(\sqrt{2}\tau_1)
\end{pmatrix}$.
After that, the relevant three-component vector
$\displaystyle\begin{pmatrix}
-i\cos\theta_1\cos(\sqrt{2}\tau_1)\cr0\cr0
\end{pmatrix}$
is turned  by $\displaystyle U^{(2)}(\tau_2-\tau_1)$ into
\be
\label{step3}
\frac{1}{2}\begin{pmatrix}
-i\big(1-\cos (2(\tau_2-\tau_1))\big)\cos\theta_1\cos(\sqrt{2}\tau_1)\cr
\sqrt{2}\sin(2(\tau_2-\tau_1))\cos\theta_1\cos(\sqrt{2}\tau_1)\cr
i\big(1+\cos(2(\tau_2-\tau_1))\big)\cos\theta_1\cos(\sqrt{2}\tau_1)
\end{pmatrix}\ .
\ee
The last component corresponds to the basis vector $|5\ra$; the second pulse $P(\theta_2)$ turns  it into the basis vector $|6\ra$  multiplying it by $\cos \theta_2$. As such it is then subjected to
$\displaystyle U^{(3)}(\tau_3-\tau_2)$.
Continuing in this way, the last non-zero entry of the final vector, that is the coefficient $\psi^{(\mu)}_N(t_f)$ which we need, reads:
\begin{eqnarray}
\nonumber
\psi^{(\mu)}_N(t_f)&=&(-1)^{K+1}\left[\prod_{i=1}^{K+1} \cos \theta_i\right] \Biggl[
\cos\left(\sqrt{2}\tau_1\right)\,
\left(\prod_{i=2}^{K+1}\,\frac{1+\cos(2(\tau_i-\tau_{i-1}))}{2} \right)\,\times\\
&&\hskip 2cm \times \cos\left(\sqrt{2}(\tau_{K+2}-\tau_{K+1})\right)\Biggr]
\equiv(-1)^{K+1}\,\chi^\theta_N\ \phi^\tau_N\ ,
\label{lastcomp}
\end{eqnarray}
where $\chi^\theta_N$ denotes the first bracket, namely the contribution from random pulses,
while $\phi^\tau_N$ that from random time intervals between pulses.
By averaging over the noise, one can then compute the mean fidelity
$\la F\ra$ and thus study the impact of the noise on the robustness of the communication line.
In the next subsections we will study in detail state transfer degradation along the chain both in
presence of independent and correlated stochastic processes. As we shall see, the transfer protocol
turns out to be more robust in the latter case, {\it i.e.} in presence of time-correlations.

\subsection{Independent noise}
The most common noise likely to affect spin chain communication
lines is that generated by un-correlated disturbances: it can be
described by totally independent stochastic variables
$\{\theta_i\}$ and $\{\tau_i\}$ with uniform distributions. In
this case, the probability density ${\cal P}(\mu)={\cal
P}(\theta,\tau)$ factorizes
\begin{eqnarray}
\label{factorized}
&&{\cal P}(\mu)={\cal P}_{pulse}(\theta)\ {\cal P}_{time}(\tau)\ ,\\
\label{factorized-pulse}
&&{\cal P}_{pulse}(\theta)={\cal P}_{pulse}(\theta_{K+1})\ {\cal P}_{pulse}(\theta_K)\ldots
{\cal P}_{pulse}(\theta_1)\ ,\\
&&{\cal P}_{time}(\tau)={\cal P}_{time}(\tau_{K+2})\ {\cal P}_{time}(\tau_{K+1})\ldots{\cal P}_{time}(\tau_1)\ ,
\end{eqnarray}
where, for simplicity, we have assumed the same probability distribution for all pulses and all time variables.
Thus, the computation of $\langle \psi^{(\mu)}_N \rangle$ and
$\langle \big|\psi^{(\mu)}_N\big|^2 \rangle$, needed in the evaluation of the fidelity,
simplifies,
\begin{equation}
\label{average-independent}
\langle \psi^{(\mu)}_N \rangle= (-1)^{K+1}\,\langle\chi^\theta_N\rangle\ \langle\phi^\tau_N\rangle\ ,\qquad
\langle \big|\psi^{(\mu)}_N\big|^2 \rangle = \langle\big(\chi^\theta_N\big)^2\rangle\
\langle\big(\phi^\tau_N\big)^2\rangle\ ,
\end{equation}
reducing~(\ref{averstate}) to the product of integrals for each stochastic variable.

In addition, we shall assume the random variables to be uniformly distributed in the intervals
$\theta_i\in[-\epsilon_\theta,\epsilon_\theta]$ and $\tau_i\in[-\epsilon_\tau, \epsilon_\tau]$,
around the perfect transfer values $\theta_i=\,0$ and $\tau_i=\,0$, so that
${\cal P}_{pulse}(\theta_i)=1/(2\epsilon_\theta)$ and ${\cal P}_{time}(\tau_i)=1/(2\epsilon_\tau)$.
However, note that, besides on $\tau_1$, the quantity $\phi^\tau_N$ in
(\ref{lastcomp}) depends also on differences of the variables $\tau_i$. Therefore, in evaluating the
averages $\langle\phi^\tau_N\rangle$ and $\langle\big(\phi^\tau_N\big)^2\rangle$, it is
convenient to introduce a new set of independent random variables,
$\delta_i\equiv\tau_i-\tau_{i-1}$, $i=2,\ldots, K+2$;
being linear combinations of two
uniformly distributed stochastic variables, these differences
obey a triangular distribution \cite{Grinstead}:
\begin{equation}
{\cal P}_{time}(\delta_i)=\frac{2\epsilon_\tau-|\delta_i|}{4\epsilon^2_\tau}\ ,\qquad
-2\epsilon_\tau\leq \delta_i \leq 2\epsilon_\tau\ .
\end{equation}
Taking this into account, the explicit computation yields:
\begin{eqnarray}
&&\langle \psi^{(\mu)}_N(t_f)\rangle= (-1)^{K+1} \left[\frac{\sin\epsilon_\theta}{\epsilon_\theta}\right]^{K+1}\
\left[\frac{\sin(\sqrt{2}\epsilon_\tau)}{\sqrt{2}\epsilon_\tau}\right]^3\
\left[\frac{1}{2}\left(1+\left(\frac{\sin 2\epsilon_\tau}{2\epsilon_\tau}\right)^2\right)\right]^K\ ,\\
\nonumber
&&\langle \big|\psi^{(\mu)}_N(t_f)\big|^2 \rangle =
\left[\frac{1}{2}\left(1+\frac{\sin2\epsilon_\theta}{2\epsilon_\theta}\right)\right]^{K+1}\
\left[\frac{1}{2}\left(1+\frac{\sin2\sqrt{2}\epsilon_\tau}{2\sqrt{2}\epsilon_\tau}\right)\right]\\
&&\hskip 2cm \times \Bigg[\frac{1}{2}\Bigg(1+
\left(\frac{\sin2\sqrt{2}\epsilon_\tau}{2\sqrt{2}\epsilon_\tau}\right)^2\Bigg)\Bigg]\
\Bigg[\frac{1}{8}\Bigg(3+4\left(\frac{\sin2\epsilon_\tau}{2\epsilon_\tau}\right)^2
+\left(\frac{\sin4\epsilon_\tau}{4\epsilon_\tau}\right)^2\Bigg)\Bigg]^K\ .
\end{eqnarray}
All terms appearing in the square brackets above are less or equal to one, becoming
smaller and smaller as $\epsilon_\theta$ and $\epsilon_\tau$, measuring the strength
of the noise, increase. Consequently, as the chain becomes large,
both $\langle \psi^{(\mu)}_N(t_f)\rangle$ and
$\langle \big|\psi^{(\mu)}_N(t_f)\big|^2 \rangle$ become small.
Therefore, in presence of uniformly distributed noise,
the fidelity $\langle F\rangle$ in (\ref{avvfid}) as function of the length $K$ of the transmission line,
decreases following a power law,
the faster the more $\epsilon_\theta$ and $\epsilon_\tau$ differ from zero (see Fig.(2)):
the advantage of using a quantum communication line with respect to
a classical one becomes then rapidly ineffective. Nevertheless, if the errors induced by the noise
are reasonably small (less than one percent), the average fidelity remains above its
corresponding classical value $F=2/3$ for quite long chains ($N>900$).
\begin{figure}[t]
\centering
\includegraphics[width=14cm,height=6cm,angle=0]{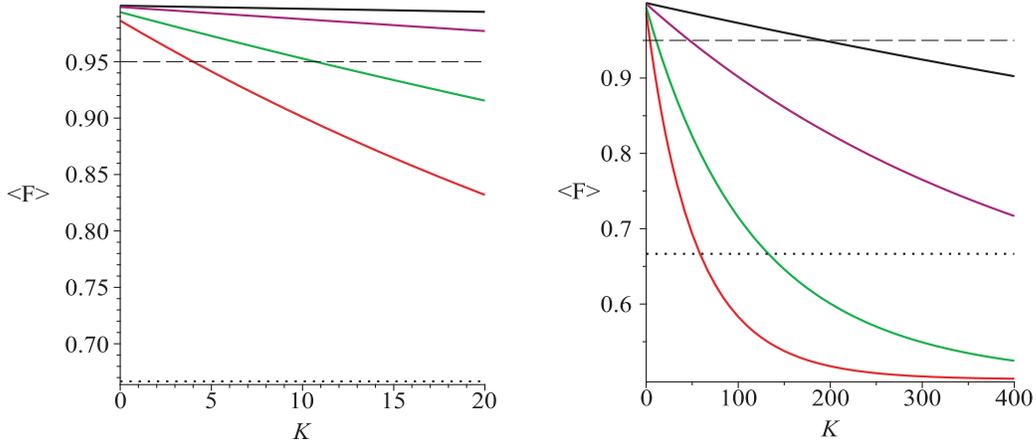}
\caption{(color online)\small Average fidelity for various values of $\epsilon_\theta=\epsilon_\tau=\epsilon$
in terms the chain length. (left figure, for a short chains of length up to $\approx 60$, right figure for long chains of length up to $\approx 1200$: $N\approx 3K$).
From top to bottom, $\epsilon=0.01,  0.02,  0.04,  $ and $0.06 $.
The dashed line represent a fidelity of 0.95 and the dotted line represents the maximum fidelity, $F=2/3$, reached using a classical communication line.
} \label{fidelity1}
\end{figure}

\subsection{Correlated noise}

The presence of correlations in the disorder affecting the spin chain
is a concrete possibility in actual experimental realizations
of the transmission protocol.
Indeed, the presence of correlations among subsequent pulses is likely to occur in practice
due to the inevitable imperfections of the apparatus which produces them, as well as in the
time intervals between pulse activations. In such instances, correlations are likely to arise,
affecting in various ways the robustness of the spin chain transmission line.
We shall concentrate on noise involving the pulses, {\it i.e.} the stochastic process $\theta$;
the disturbances affecting the time-intervals between pulses, described by
the process $\tau$, can be similarly treated and will be briefly discussed at the end
of the section.%
\footnote{Also here, for simplicity, the two processes $\theta$ and $\tau$ are assumed to be independent;
correlations between pulses and time intervals are surely possible in principle,
but certainly less likely than those between $\theta$'s and $\tau$'s variables themselves.}

In presence of correlations among pulses, the probability density ${\cal P}_{pulse}(\theta)$
can no longer be written as the product of independent probabilities as in (\ref{factorized-pulse}).
The simplest generalization of (\ref{factorized-pulse}) involves joint probabilities with one-step correlations
based on the conditional probabilities
\begin{equation}
\label{joint-probability}
{\cal P}_{pulse}(\theta_{i+1} | \theta_i)= q\, \delta(\theta_{i+1}-\theta_i)\, +\, (1-q)\, {\cal P}_{pulse}(\theta_{i+1})\ ,
\end{equation}
specifying the probability that the stochastic variable corresponding to the $i+1$-th pulse take the value $\theta_{i+1}$ conditioned on the stochastic
variable corresponding to the preceding pulse having taken the value $\theta_i$, this being valid for all $i=1,2,\ldots,K$.
The parameter $q\in [0,1]$ measures the amount of correlations between the
stochastic variables $\theta_{i+1}$ and $\theta_i$, which is maximal for $q=1$, while,
for vanishing $q$, $\theta_{i+1}$ and $\theta_i$ are independent stochastic variables.
Then, we shall assume the stochastic process $\theta=\{\theta_i\}$ to be characterized by one-step correlations, namely that its
joint probability distributions are of the form
\begin{equation}
\label{correlated-prob}
{\cal P}_{pulse}(\theta_{K+1},\cdots,\theta_{2},\theta_1)=
{\cal P}_{pulse}(\theta_{K+1}|\theta_K)\ldots {\cal P}_{pulse}(\theta_2 | \theta_1)\
{\cal P}_{pulse}(\theta_1)\ .
\end{equation}
In determining the fidelity in (\ref{avvfid}), one now needs to use this expression
in computing the averages over the noise.
In practice, since the noise affects only pulses and not the time-intervals between them,
the quantity~(\ref{lastcomp}) which enters~(\ref{averstate}) together with its square modulus, reduces to $(-1)^{K+1}\chi^\theta_N$, thus one needs evaluate integrals of the form
\begin{equation}
\label{correlated-average}
{\cal I}=\int d^{K+1}\theta\ f(\theta_{K+1})\,
{\cal P}_{pulse}(\theta_{K+1}|\theta_K) \ldots f(\theta_2)\, {\cal P}_{pulse}(\theta_2 | \theta_1)\,
f(\theta_1)\, {\cal P}_{pulse}(\theta_1)\ ,
\end{equation}
where $f(\theta)$ is either $\cos\theta$ or $\cos^2\theta$.

In order to estimate the effects of correlated noise on the behavior of the averaged fidelity, we shall
assume that the stochastic variables $\theta_i$ take only three possible values,
$-\epsilon_\theta$, $0$ and $\epsilon_\theta$,
with a probability distribution given by $(0\leq p\leq1/2)$:
\begin{equation}
\label{corr-pulse}
{\cal P}_{pulse}(\theta_i)=
\left\lbrace
\begin{array}{l}
p\hskip 1.6cm\ \theta_i =\pm \epsilon_\theta\ ,\\
\\
1-2p\ \ \ \ \ \ \theta_i =0\ .\\
\end{array}\right.
\end{equation}
Then, the Dirac delta in (\ref{joint-probability}) becomes a Kronecker delta and the integrals of the form (\ref{correlated-average})
reduce to sums: $\int d^{K+1}\theta \to \sum_{\theta_1\ldots\theta_{K+1}}$ and
can be cast in a compact form by adopting a transition matrix formalism.
That is, one introduces the orthonormal vectors
\begin{equation}
|-\epsilon_\theta\ra=\left(\begin{array}{c} 0\\ 0\\ 1\end{array}\right)\ ,\h
|0\ra=\left(\begin{array}{c} 0\\ 1\\ 0\end{array}\right)\ ,\h
|\epsilon_\theta\ra=\left(\begin{array}{c} 1\\ 0\\ 0\end{array}\right)\ ,
\end{equation}
 a probability vector $|{\cal P}_{pulse}\rangle$
with components $\langle\theta|{\cal P}_{pulse}\rangle={\cal P}_{pulse}(\theta)$, and collect the conditional
probabilities into a $3\times 3$ transition matrix with entries
$\langle\theta | \mathbb{P}_{pulse} |\theta'\rangle= {\cal P}_{pulse}(\theta | \theta')$.
Explicitly, using (\ref{joint-probability}) and (\ref{corr-pulse}), one finds:
\begin{eqnarray}
\label{p-matrix}
\mathbb{P}_{pulse}&=&\begin{pmatrix}
q+(1-q)p&(1-q)p&(1-q)p\cr
(1-q)(1-2p)&q+(1-q)(1-2p)&(1-q)(1-2p)\cr
(1-q)p&(1-q)p&q+(1-q)p\cr
\end{pmatrix}\ ,\\
\label{vector-p}
|{\cal P}_{pulse}\rangle&=&\left(\begin{array}{c} p\\ 1-2p\\ p\end{array}\right)\ .
\end{eqnarray}
Further, by introducing the diagonal $3\times 3$ matrix
\begin{equation}
\label{diagaver} \mathbb{F}=\begin{pmatrix}
f(\epsilon_\theta)&&\cr &f(0)&\cr &&f(-\epsilon_\theta)\cr
\end{pmatrix}\ ,
\end{equation}
the average (\ref{correlated-average}) can be formally rewritten as the following matrix element:
\begin{equation}
\label{matrix-form} {\cal I}=\langle\Theta|
\big(\mathbb{F}\,\mathbb{P}_{pulse}\big)^K\, \mathbb{F}\, |{\cal
P}_{pulse}\rangle \ ,
\end{equation}
where the final vector $\langle\Theta|$ is the sum of the three basis vectors, explicitly given by
\begin{equation}
\label{vector-theta}
\langle\Theta|=\sum_{\theta=\{\pm\epsilon_\theta,0\}} \langle \theta|=(1\ 1\ 1)\ .
\end{equation}
By recalling that $f(\epsilon_\theta)$ is either $\cos\epsilon_\theta$ or its square,
this result allows evaluating for any fixed $K$ the effects of correlated pulse noise
in the fidelity (\ref{avvfid}) as a function of the parameters $q$ and $p$ (see Fig.(3)).
As expected, the efficiency of the qubit transfer through the chain degrades in presence of
the correlated noise, but in a less dramatic way if compared with its behaviour in presence
of disturbances with no correlations;
indeed, almost perfect transfer is achieved for high correlated noise
even when the error parameter $p$, the probability
for the stochastic variables $\theta_i$ to differ from
the perfect transfer value $\theta_i=\,0$, is as large as 1/2.

\begin{figure}[t]
\centering
\includegraphics[width=8cm,height=8cm,angle=0]{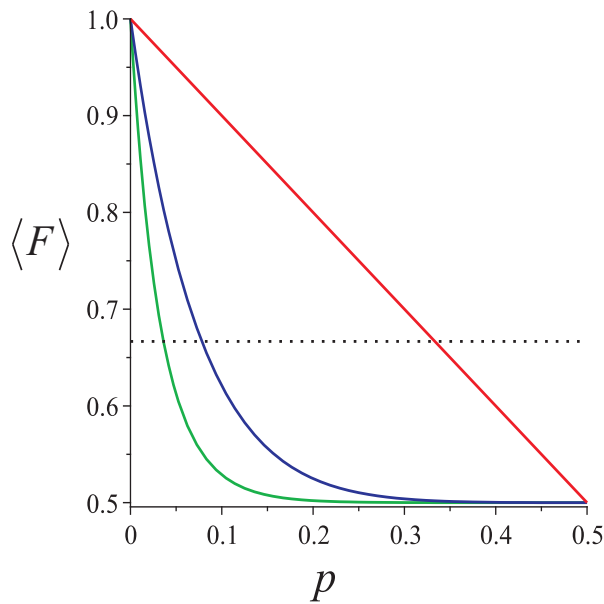}
\caption{(color online)\small Average fidelity for various values of $q$ in terms of the error probability $p$.
From bottom to top, $q=0$  (no correlation), $0.9 $ and $1$ (full correlation). For all the curves $\epsilon=0.5$ and $K=100$.
} \label{fidelity2}
\end{figure}

In particular, for $q=1$, {\it i.e.} when the correlations between subsequent
pulses are maximal, one finds:
\begin{equation}
\langle F\rangle= 1-p +\frac{p}{3}\Big[2\, (\cos\epsilon_\theta)^{K+1}
+(\cos\epsilon_\theta)^{2K+2}\Big]\ ,
\end{equation}
clearly showing that external stochastic noise containing
correlations, hence some sort of correlations, disturbs in a milder way
the spin chain transmission protocol. Specifically, for long chains,
as $K$ becomes large, the averaged fidelity reaches the asymptotic value
\begin{equation}
\langle F\rangle\sim 1-p \ ,
\end{equation}
which can still be close to unity, provided $p$ is sufficiently small.

This result has to be compared with the one obtained in the case $q=\,0$, when no correlations
are present and all stochastic variables $\theta_i$ are independent. Also in this case, the pulse
noise contribution to the averaged fidelity can be exactly computed, yielding:
\begin{equation}
\langle F\rangle= \frac{1}{2}+\frac{1}{6}\Big[1-4p\sin^2(\epsilon_\theta/2)\Big]^{K+1}
+\frac{1}{3}\Big[1-2p\sin^2\epsilon_\theta)\Big]^{K+1}\ .
\end{equation}
Since $0\leq p\leq 1/2$, the square brackets above are always $\leq1$, so that as $K$ increases
the fidelity rapidly approaches its asymptotic value of 1/2; this is precisely the behaviour
encountered in the previous section while discussing independent noise.

Similar results are obtained when correlations are present in the
stochastic variables $\tau_i$, affecting the time intervals
between the pulses: the joint probabilities ${\cal
P}_{time}(\tau_i|\tau_j)$ can be taken as in
(\ref{joint-probability}). By assuming that the stochastic
variables $\tau_i$ take only the three possible values
$-\epsilon_\tau$, $0$ and $\epsilon_\tau$, with a probability
distribution ${\cal P}_{time}(\tau_i)$ similar to the one in
(\ref{corr-pulse}), the computation of the noise contributions to
the averaged fidelity can be treated as in the previous case,
leading to contributions of the form (\ref{matrix-form}).
However, note that now, except for the first contribution, the
matrix $\mathbb{F}$ is no longer diagonal since it involves time
differences; indeed, instead of $\mathbb{F}$ in (\ref{diagaver}),
one has to use one with entries: $\langle\tau |
\mathbb{F}|\tau'\rangle= f(\tau-\tau')$, where, recalling
(\ref{lastcomp}), $f(\epsilon_\tau)$ is either $(1+\cos
2\epsilon_\tau)/2$, or its square.

The behavior of the averaged fidelity in terms of the probability $p$, for different values
of the correlation parameter $q$, at fixed $K$ and $\epsilon_\tau$, is qualitatively similar
to the one discussed before in the case of correlated pulses, given in Fig.(3).
In particular, also in this case one observes that the fidelity is less affected by
the presence of correlated noise, to the extent that when $q=1$ it acquires a constant
value, independent from the length of the chain:
\begin{equation}
\la F\ra=1-\frac{2p}{3}\sin^2 2\epsilon_\tau\, \big(2+\sin^2 2\epsilon_\tau\big)\ .
\end{equation}
This result is easily understandable; indeed, when $q=1$, all intermediate three-site sub-chains
(lower picture in Fig.(1)) remain perfect state transfer chains even in presence of noise,
and only the first and the last two-site sub-chain fail to transfer the state perfectly, so that
the actual length of the transmission line becomes effectively irrelevant.

\section{Discussion}

We have studied the effect of imperfections in the external
control in  schemes for perfect transmission of quantum states
through a quasi-dimensional chain \cite{Pemberton}. Such chains
are to be connected to each other to form larger two and three
dimensional networks \cite{Pemberton},\cite{KSA} through which
qubit states are to be routed from any point to any other point
through the natural dynamics of the underlying XY Hamiltonia when
assisted by global control pulses. These protocols are  by
construction robust against known localized imperfections in the
network structure (un-desired couplings, etc). Our study shows
that as long as the quasi-one dimensional chains is concerned,
such schemes are also robust against imprecision in the sequence
of applied global pulses, at least for moderate level of noise
and for moderate lengths of the chain.   Remarkably we have found
that when the noise in successive applications of the pulses are
correlated, the efficiency of the protocol is less damaged
compared with the case when there is no correlations. Despite the
complications of natural dynamics intervened by global external
pulses, we have been able to derive exact expressions for the
fidelity of state transfer, by taking advantage of the sequential
dynamics of the many-body state and following only the evolution
of the relevant coefficient which is necessary for calculation of
the fidelity.

\section*{Acknowledgements:}

V. K. would like to thank The Abdus Salam ICTP for its hospitality during the summer 2012,
where most of this paper has been prepared.

\end{document}